\documentclass[draftcls, onecolumn, 12pt]{IEEEtran}
\usepackage{amsmath,amssymb,threeparttable,placeins,enumerate,caption}
\usepackage{mathtools,relsize}
\usepackage[pdftex]{graphicx}
\usepackage{epstopdf}
\usepackage{color}
\usepackage[usenames]{xcolor}
\usepackage{amsfonts}
\usepackage{latexsym}
\usepackage{subfigure,multirow,makecell,stfloats}

\usepackage{algorithm}
\usepackage[noend]{algpseudocode}
\def\qi#1 {\fbox {\footnote {\ }}\ \footnotetext { From Qi: {\color{red}#1}}}

\newtheorem{theorem}{{{\textit{Theorem}}}}

\newtheorem{lemma}{{{\textit{Lemma}}}}
\newtheorem{corollary}{{{{\textit{Corollary}}}}}

\newtheorem{definition}{{{\textit{Definition}}}}

\newtheorem{remark}{{{\textit{Remark}}}}

\newtheorem{example}{{{\textit{Example}}}}

\newtheorem{construction}{{{\textit{Construction}}}}
\begin{document}

\title{New Construction of Optimal Type-II Binary Z-Complementary Pairs}

\author{Zhi~Gu,~\IEEEmembership{Student Member,~IEEE,}
        Zhengchun~Zhou,~\IEEEmembership{Member,~IEEE,}
        Qi~Wang,~\IEEEmembership{Member,~IEEE,}
        and~Pingzhi~Fan,~\IEEEmembership{Fellow,~IEEE}
\thanks{Z. Gu is with the School of Information Science and Technology, Southwest Jiaotong University, Chengdu, 611756, China. E-mail: goods@my.swjtu.edu.cn.}
\thanks{Z. Zhou is with the School of Mathematics, Southwest Jiaotong University, Chengdu, 611756, China. E-mail: zzc@swjtu.edu.cn.}
\thanks{Q. Wang is with the Department of Computer Science and Engineering, Southern University of Science and Technology, Shenzhen 518055, China. E-mail: wangqi@sustech.edu.cn.}
\thanks{P. Fan is with the Institute of Mobile Communications, Southwest Jiaotong University, Chengdu, 611756, China. E-mail: pzfan@swjtu.edu.cn.}
}
\maketitle

\date{}

\begin{abstract}
	
  A pair of sequences is called a Z-complementary pair (ZCP) if it has zero aperiodic autocorrelation sums at each of the non-zero time-shifts within {a} certain region,  called the zero correlation zone (ZCZ). ZCPs are categorised into two types{:} Type-I ZCPs and Type-II ZCPs. Type-I ZCPs have {the} ZCZ around the in-phase position and Type-II ZCPs have the ZCZ around the end-shift position. {Till now only a few} constructions of Type-II ZCPs are reported {in the literature}, and all {have} lengths of the form $2^m\pm1$ or $N+1$ where $N=2^a 10^b 26^c$ and $a,~b,~c$ are non-negative integers. In this paper, we {propose} a recursive construction of ZCPs based on concatenation of sequences. Inspired by Turyn's construction of Golay complementary pairs, we also propose a construction of Type-II ZCPs from known ones. The proposed constructions can generate optimal Type-II ZCPs with new flexible parameters and Z-optimal Type-II ZCPs with any odd length. In addition, we give upper bounds for the PMEPR of the proposed ZCPs. It turns out that our constructions lead to ZCPs with low PMEPR.
	
\end{abstract}


\section{Introduction}
\IEEEPARstart{A}{ PAIR} of sequences is called a Golay complementary pair (GCP), if their aperiodic autocorrelation sums (AACSs) are zero everywhere, except at the zero shift \cite{Golay51}, \cite{Golay61}. GCPs were first introduced by {Golay} in 1961 in the context of an optical problem in multislit spectrometry. Since then, {GCPs have} found extensive engineering applications for its ideal correlation properties. For example, GCPs are useful in inter-symbol interference channel estimation \cite{Spasojevic01,Wang07}, radar waveform designs \cite{Budisin91,Pezeshki08,Kumari18}, asynchronous multi-carrier code-division multiple access (MC-CDMA) communications \cite{Chen01,Liu15}, and peak-to-mean envelope power ratio (PMEPR) control in multi-carrier communications \cite{Davis99,Wang14}.

The main drawback of the GCPs {is their} limited availability for various lengths. It {was} conjectured that binary GCPs are available only for lengths of the form $2^a 10^b 26^c$ where $a,~b,~c$ are non-negative integers \cite{Borwein04}. By computer search{,} the conjecture {has} been verified for binary GCPs of length up to $100$ \cite{Borwein04}. In search of binary sequence pairs of other lengths, {Fan, Yuan and Tu~\cite{Fan07}} proposed
{the concept of} Z-complementary pair (ZCP) in 2007, {which is a pair of sequences whose aperiodic autocorrelation sums are zero not everywhere but within a certain region called {\em zero correlation zone} (ZCZ)}. Based on their lengths, binary ZCPs {are} categorised into two types: odd-length (OB-ZCPs) and even-length ZCPs (EB-ZCPs). It was {further} conjectured by Fan, Yuan and Tu~\cite{Fan07} that {``\textit{For OB-ZCPs, the maximum zero correlation zone is given by $Z_{\max}=(N+1)/2$, and for EB-ZCPs, given that the lengths $N \neq 2^a 10^b 26^c$, the ZCZ is upper bounded by $N-2$.}"}
In 2011 Li \textit{et al.}~\cite{li2010} proved the conjecture for OB-ZCPs. However, a systematic construction of OB-ZCPs was still {unknown}.

In 2014, Liu, Udaya and Guan~\cite{Liu14-2} made remarkable progress towards this open problem and proposed a systematic construction of {OB-ZCPs. The generated optimal} sequence pairs achieve the maximum ZCZ {of width} $(N+1)/2$ {as well as the} minimum AACSs magnitude of $2$ at each time-shift outside the ZCZ \cite{Liu14-2}. The construction of optimal OB-ZCPs in \cite{Liu14-2} was {given by} applying the insertion method on binary Golay-Davis-Jedwab (GDJ) sequences. In 2014, Liu, Udaya and Guan~\cite{Liu14} {also confirmed the conjecture for EB-ZCPs} that $Z_{\max}\leq N-2$.
{Recently, a lot of work has been done for constructing EB-ZCPs (for example, see~\cite{Chen17,Adhikary18,Xie18}).} Besides the lengths, in \cite{Liu14-2} {Liu, Udaya and Guan further categorised ZCPs based on their correlation properties: Type-I ZCPs are sequence pairs having zero AACSs at each time-shift within the ZCZ around the in-phase position, while Type-II ZCPs are those having zero AACSs at each time-shift within the ZCZ around the end-shift position.} Type-I ZCPs can be effectively used in quasi-synchronous CDMA (QS-CDMA) systems \cite{Suehiro94,Long98,Fan99,Tang06,Tang10,Lee10,Lee98}, which are tolerant of small-signal arrival delays. On the other hand, Type-II ZCPs are useful in wide-band wireless communication systems where the minimum interfering-signal delay can assume a large value. This is because the ZCZ of Type-II ZCP is designed for large time-shifts, and thus the asynchronous interfering signals arriving at the receiver after large delays can be rejected. A typical example of such a channel with large delays is the sparsely populated rural and mountainous areas \cite{Lee10}. {{In some other important applications, like designing preamble sequences in OFDMA systems \cite{IEEE}, where PMEPR plays a very important role, Type-II ZCPs may be advantageous over Type-I ZCPs because of its huge availability with flexible lengths.
}
Moreover, Type-I and Type-II ZCPs can also be used to construct complementary sets \cite{Avik cs} and Z-complementary sets \cite{Avik zccs}.}

Till now there are only a few constructions of Type-I and Type-II constructions reported in the literature~\cite{Liu14-2,Liu14,Chen17,Adhikary18,Xie18,Avik tit,Avik iwsda}. Note that most of the constructions are based on GCPs and {thereby} have lengths of the form of $2^a+2^v$. Recently, {based on generalized Boolean functions,} Chen \cite{Chen17} gave a direct construction of those Type-I ZCPs having lengths of the form $2^{m-1}+2^v$ {and} a ZCZ {of width} $2^{\pi(v+1)-1}+2^v$, where $\pi$ is a permutation over $\{1,2,\dots,m-1\}$. Adhikary \textit{et al.} \cite{Avik tit,Avik iwsda} made further progress towards this problem and proposed a systematic construction of Type-I and Type-II ZCPs of lengths of the form $2^a 10^b 26^c+1$, by applying the insertion method on binary GCPs.
{Very recently, Shen \textit{et al.} \cite{Shen} constructed Type-II ZCP of length $2^m+3$, by inserting 3 elements into GDJ sequences.
}
An overview of known Type-II binary ZCPs is given in Table \ref{table 11}, together with their corresponding ZCZ width. For the definitions of Z-optimality and optimality of binary ZCPs, see~Definitions~\ref{def-zopt} and~\ref{def-opt}, respectively.

\begin{table*}[h]\caption{Existing and proposed binary Type-II ZCPs.}\label{table 11}
\centering
\renewcommand{\arraystretch}{1.3}
\resizebox{0.8\textwidth}{!}{
		\begin{tabular}{|c|c|c|c|c|c|}
			\hline
			Ref. & Length & ZCZ width & \makecell{Magnitude\\ outside\\ the ZCZ ($v$)}& \makecell{Remarks on \\Z-optimality}& \makecell{Remarks on \\optimality}   \\ \hline \hline
			\cite{Liu14-2}&$2^m+1$ & $2^{m-1}+1$ & $2$&Z-optimal& Optimal  \\
			\hline
			
			\cite{Liu14-2}&$2^m-1$ & $2^{m-1}$ &$2$& Z-optimal& Optimal  \\
			\hline
			
			\cite{Avik tit}&\makecell{$2^a10^b26^c+1,$\\$~a\geq1$} & $2^{a-1}10^b26^c+1$& $2$ & Z-optimal& Optimal  \\
			\hline
			
			\cite{Avik tit}&\makecell{$10^b+1,$\\$~b\geq1$} & $4\times 10^{b-1}+1$ &$2$& Not Z-optimal& Not optimal  \\
			\hline
			
			\cite{Avik tit}&\makecell{$26^c+1,$\\$~c\geq1$} & $12\times 26^{c-1}+1$ &$2$& Not Z-optimal&Not optimal  \\
			\hline
			\cite{Avik tit}&\makecell{$10^b26^c+1,$\\$~b,~c\geq1$} & $12\times 10^b 26^{c-1}+1$&$2$ & Not Z-optimal& Not optimal  \\
			\hline
            {\cite{Shen}} & {$2^m+3$} & $2^{m-1}+2$ & $v\in\{2,6\}$ & Z-optimal& Not optimal\\
            \hline
			Theorem \ref{Theorem 1}&\makecell{$2N+1,$\\$N\in \mathbb{Z}^+$ } & $N+1$&$2\leq v \leq 2(2N-1)$ & Z-optimal& {\makecell{Optimal when\\ $|\rho_\mathbf{a}(\tau)+\rho_\mathbf{b}(\tau)|=1$}}  \\
			\hline
			Remark \ref{rem11}&\makecell{$2N-1,$\\$N\in \mathbb{Z}^+$ } & $N$&$2\leq v \leq 2(2N-3)$ & Z-optimal& Not optimal  \\
			\hline
			Theorem \ref{Theorem 2}&\makecell{$2^kN+2^{k-1},$\\$N\in \mathbb{Z}^+,~k\geq 2$ } & $2^kN+2^{k-1}-N$&$4\leq v \leq 2^k(2N-1)$ & \makecell{Z-optimal\\when $N=1$}& Not optimal  \\
			\hline
			{Theorem \ref{Turyn}} &\makecell{$3N,$\\$N= 2^a10^b26^c$ } & $3N-1$&$2N$ & Z-optimal& \makecell{Optimal\\when $N=1,2$}  \\
			\hline
			{Theorem \ref{Turyn}} &\makecell{$14N,$\\$N= 2^a10^b26^c$ } & $14N-1$&$4N$ & Z-optimal& \makecell{Optimal\\when $N=1$}  \\
			\hline
			Theorem \ref{Construct TCP1} &\makecell{$2N-1,$\\$N= 2^a10^b26^c$ } & $N$&$2$ & Z-optimal& Optimal  \\
			\hline
			Theorem \ref{Construct TCP2} &\makecell{$2N+1,$\\$N= 2^a10^b26^c$ } & $N+1$&$2$ & Z-optimal& Optimal  \\
			\hline
		\end{tabular}}
\end{table*}

To the best of {our} knowledge, the maximum ZCZ width for binary Type-II ZCPs of lengths of the form $10^b+1$, $26^c+1$ and $10^b26^c+1$ are $4\times 10^{b-1}+1$, $12\times 26^{c-1}+1$ and $12\times 10^b26^{c-1}+1$, respectively. {Furthermore, there is no construction of Type-II EB-ZCPs in the literature}.

Motivated by the constructions reported in \cite{Liu14-2,Avik tit}, in search of ZCPs with larger ZCZ widths, we propose an iterative construction of Type-II binary ZCPs of {both even and odd lengths}. Our proposed construction can generate Z-optimal Type-II OB-ZCPs having lengths of any odd length, {and also} optimal Type-II EB-ZCPs for certain cases. In fact, our proposed construction can generate Type-II ZCPs with more flexible lengths which {were unknown before}. As a comparison with {previous results}, our results are given in Table \ref{table 11}. We {further} list down the ``best possible" ZCPs up to length $30$ in Table \ref{ref2}. {The} term ``best possible", {means that the ZCPs have} the closest possible autocorrelation properties to {those} of the optimal Type-II ZCPs. Note that the sequence pairs, whose lengths are given in bold letters in Table \ref{ref2}, were not reported before.

{The rest of this paper} is organized as follows.
In Section \ref{section 2}, {we introduce some basic definitions and preliminary results about ZCPs, and the peak-to-mean envelope power ratio (PMEPR) control problem in code-keying MC communications.}
In Section \ref{section 3}, we propose {a generic} construction of Type-II ZCPs, {which allows generating both OB-ZCPs and EB-ZCPs.}
In addition, we propose a construction of optimal Type-II OB-ZCPs in Section \ref{section 4}.
{In Section \ref{section 5}, we analyse the PMEPR  of the proposed ZCPs.
}
Finally, Section~\ref{section 6} concludes the paper by some future work.

\begin{table}[ht!]\caption{``Best possible" Type-II sequence pairs of length up to $30$}\label{ref2}
	\renewcommand{\arraystretch}{0.78}
	\resizebox{\columnwidth}{!}{
		\begin{tabular}{|c|c|c|c|}
			\hline
			$N$ & Type & $\left(
			\begin{array}{c}
			\mathbf{a} \\
			\mathbf{b} \\
			\end{array}
			\right)
			$ & $\left|\rho_\mathbf{a}(\tau)+\rho_\mathbf{b}(\tau)\right|_{\tau=0}^{N-1}$ \\ \hline \hline
			
		2 & GCP & $\bigg( \begin{matrix} ++ \\ +- \end{matrix} \bigg)$ & (4,0) \\
		\hline
		3 & Optimal OB-ZCP & $\bigg( \begin{matrix} +++ \\ ++- \end{matrix} \bigg)$ & (6,2,0) \\
		\hline
		4 & GCP & $\bigg( \begin{matrix} +++- \\ ++-+ \end{matrix} \bigg)$ & $(8,\mathbf{0}_{3})$ \\
		\hline
		5 & Optimal OB-ZCP & $\bigg( \begin{matrix} ---++ \\ --+-- \end{matrix} \bigg)$ & $(10,\mathbf{2}_{2},\mathbf{0}_{2})$ \\
		\hline
		\textbf{6} & Optimal EB-ZCP & $\bigg( \begin{matrix} ++++-- \\ +++-++ \end{matrix} \bigg)$ & $(12,4,\mathbf{0}_{4})$ \\
		\hline
		7 & Optimal OB-ZCP & $\bigg( \begin{matrix} --+-+-- \\ --++-++ \end{matrix} \bigg)$ & $(14,\mathbf{2}_{3},\mathbf{0}_{3})$ \\
		\hline
		8 & GCP & $\bigg( \begin{matrix} +++-++-+ \\ +++---+- \end{matrix} \bigg)$ & $(16,\mathbf{0}_{7})$ \\
		\hline
		9 & Optimal OB-ZCP & $\bigg( \begin{matrix} +-+++++-- \\ +-++---++ \end{matrix} \bigg)$ & $(18,\mathbf{2}_{4},\mathbf{0}_{4})$ \\
		\hline
		10 & GCP & $\bigg( \begin{matrix} ++-+-+--++ \\ ++-+++++-- \end{matrix} \bigg)$ & $(20,\mathbf{0}_{9})$ \\
		\hline
		
		\textbf{12} & Z-optimal EB-ZCP & $\bigg( \begin{matrix} ++++--+++-++ \\ ++++-----+-- \end{matrix} \bigg)$ & $(24,8,\mathbf{0}_{10})$ \\
		\hline
		
		\textbf{14} & EB-ZCP & $\bigg( \begin{matrix} --+-+----++-++ \\ --+-+--++--+-- \end{matrix} \bigg)$ & $(28,\mathbf{4}_{3},\mathbf{0}_{10})$ \\
		\hline
		15 & Optimal OB-ZCP & $\bigg( \begin{matrix} -++-+++++-+-+++ \\ -++-+++--+-+--- \end{matrix} \bigg)$ & $(30,\mathbf{2}_7,\mathbf{0}_{7})$ \\
		\hline
		16 & GCP & $\bigg( \begin{matrix} +++-++-++++---+- \\ +++-++-+---+++-+ \end{matrix} \bigg)$ & $(32,\mathbf{0}_{15})$ \\
		\hline
		17 & Optimal OB-ZCP & $\bigg( \begin{matrix} -++++-+-+-++--+++ \\ -++++-+--+--++--- \end{matrix} \bigg)$ & $(34,\mathbf{2}_{8},\mathbf{0}_{8})$ \\
		\hline
		\textbf{18} & EB-ZCP & $\bigg( \begin{matrix} +-+++++--+-++---++ \\ +-+++++---+--+++-- \end{matrix} \bigg)$ & $(36,\mathbf{4}_{4},\mathbf{0}_{13})$ \\
		\hline
		\textbf{19} & Optimal OB-ZCP & $\bigg( \begin{matrix} +-+++++--++--+-+-++ \\ +-+++++----++-+-+-- \end{matrix} \bigg)$ & $(22,\mathbf{2}_{9},\mathbf{0}_{9})$ \\
		\hline
		20 & GCP & $\bigg( \begin{matrix} ++-+-+--++++-+++++-- \\ ++-+-+--++--+-----++ \end{matrix} \bigg)$ & $(20,\mathbf{0}_{9})$ \\
		\hline
		
		\textbf{21} & Optimal OB-ZCP & $\bigg( \begin{matrix} -++-+-+++-  -++++++--++ \\ -++-+-+++-  +------++-- \end{matrix} \bigg)$ & $(42,\mathbf{2}_{10},\mathbf{0}_{10})$ \\
		\hline
		
		\textbf{24} & Z-optimal EB-ZCP & $\bigg( \begin{matrix} ++++--+++-++++++-----+-- \\ ++++--+++-++----+++++-++ \end{matrix} \bigg)$ & $(48,16,\mathbf{0}_{22})$ \\
		\hline
		
		26 & GCP & $\bigg( \begin{matrix} ++++-++--+-+-+--+-+++--+++ \\ ++++-++--+-+++++-+---++--- \end{matrix} \bigg)$ & $(52,\mathbf{0}_{25})$ \\
		\hline
		
		\textbf{28} & Z-optimal EB-ZCP & $\bigg( \begin{matrix} --+-+----++-++--+-+--++--+-- \\ --+-+----++-++++-+-++--++-++ \end{matrix} \bigg)$ & $(56,\mathbf{8}_{3},\mathbf{0}_{24})$ \\
		\hline
		
		\textbf{30} &{ Z-optimal} EB-ZCP & {$\bigg( \begin{matrix} +-++-+-+-++--+-++--+---+++-++- \\
		+-++-+-+-++-+-+++-+-+++---+--+ \end{matrix} \bigg)$} & {$(60,20,\mathbf{0}_{28})$} \\
			\hline
		\end{tabular}}
\end{table}

\section{Preliminaries}\label{section 2}

In this section, we recall some definitions and bounds of binary ZCPs. Before that, {we fix} some notations which will be used throughout the paper.

\begin{itemize}
	\item $+$ and $-$ denote  $1$ and $-1$, respectively.
	\item $\mathbf{0}_L$ and $\mathbf{1}_L$  denote length-L vectors whose elements are all $0$ and $1$, respectively.
	\item $\overleftarrow{\mathbf{c}}=(c_{N_1-1},c_{N_1-2},\cdots,c_0)$ denotes the reverse of sequence {$\mathbf{c} = (c_0, \ldots, c_{N_1 - 2}, c_{N_1 - 1}) $}.
\item $\mathbf{c}\mid \mid\mathbf{d}$ denote the horizontal concatenation of sequences $\mathbf{c}$ and $\mathbf{d}$.
	\item $\mathbf{c} \otimes \mathbf{d}$ denotes the Kroneker product of the sequences $\mathbf{c}$ and $\mathbf{d}$ of lengths $N_1$ and $N_2$, respectively, {i.e.,}
	\begin{equation*}
\mathbf{c} \otimes \mathbf{d}=(c_0d_0,c_0d_1,\cdots,c_0d_{N_2-1},\cdots, c_{N_1-1}d_0,c_{N_1-1}d_1,\cdots,c_{N_1-1}d_{N_2-1}). 
	\end{equation*}
\end{itemize}

{In the following, we first give the definition of aperiodic correlation, and then define the deletion function.}

\begin{definition}
  For two length-$N$ binary sequences $\mathbf{c}$ and $\mathbf{d}$, their {\em aperiodic cross-correlation function} is defined as
 \begin{equation}\label{Definition-acf}
 {
   \rho_{\mathbf{c,d}}(\tau)=
    \left\{
      \begin{array}{ll}
        \sum_{k=0}^{N-1-\tau}c_kd_{k+\tau}, & 0\leq\tau\leq N-1, \\
        \sum_{k=0}^{N-1-\tau}c_{k+\tau}d_k, & -(N-1)\leq\tau\leq-1, \\
        0, & |\tau|\geq N.
      \end{array}
    \right.}
 \end{equation}
 When $\mathbf{c} = \mathbf{d}$, {the function $\rho_{\mathbf{c,d}}(\tau)$ is called the {\em aperiodic autocorrelation function}} (AACF) of $\mathbf{c}$, denoted {by} $\rho_\mathbf{c}(\tau)$ for simplicity.
\end{definition}

\begin{definition}\label{}(Deletion Function)
For a sequence $\mathbf{c}=(c_0,c_1,\ldots,c_{N-1})$ and an integer $r\in\{0,1,\ldots,N-1\}$, define $\mathcal{V}(\mathbf{c},r)$ as a deletion  function of $\mathbf{c}$ as
\begin{equation*}
\mathcal{V}(\mathbf{c},r)
=\begin{cases}
(c_1,c_2,\ldots,c_{N-1}), & \hbox{ if }r=0; \\
(c_0,c_1,\ldots,c_{N-2}), & \hbox{ if }r=N-1; \\
(c_0,c_1,\ldots,c_{r-1},c_{r+1},\ldots,c_{N-1}), & \hbox{otherwise,}
\end{cases}
\end{equation*}
where $r$ denotes the deletion position.
\end{definition}

{In what follows, we give a series of definitions on a pair of sequences with desirable aperiodic autocorrelation sums, and certain bounds on these sequence pairs.}

\begin{definition}\label{Definition-GCP}
  A pair of sequences $\mathbf{c}$ and $\mathbf{d}$ of length $N$ is called a {\em Golay complementary pair} (GCP) if and only if
\begin{equation*}
  \rho_\mathbf{c}(\tau)+\rho_\mathbf{d}(\tau)=0,
\end{equation*}
{for all $1 \leq \tau \leq N -1$.}
\end{definition}

\begin{definition}\label{Definition-Type-I ZCP}(Type-I binary ZCPs)
  A pair of binary sequences $\mathbf{c}$ and $\mathbf{d}$ of length $N$ is called a {Type-I Z-complementary pair} (ZCP) with ZCZ {of width} $Z$, if and only if
\begin{equation*}
  \rho_\mathbf{c}(\tau)+\rho_\mathbf{d}(\tau)=0,
\end{equation*}
for all $1\leq\tau\leq Z-1$, and $\rho_\mathbf{c}(Z)+\rho_\mathbf{d}(Z)\neq0$.
\end{definition}

\begin{definition}\label{Definition-Type-II ZCP}(Type-II binary ZCPs)
A pair of binary sequences $\mathbf{c}$ and $\mathbf{d}$ of length $N$ is called a Type-II ZCP with ZCZ of width $Z$, if and only if
\begin{equation}
  \rho_\mathbf{c}(\tau)+\rho_\mathbf{d}(\tau)=0, \text{ for all } N-Z+1\leq\tau\leq N-1,
\end{equation}
and $\rho_\mathbf{c}(N-Z)+\rho_\mathbf{d}(N-Z)\neq0$.
\end{definition}

Clearly, when $Z = N$, both Type-I and Type-II ZCPs become GCPs.

%
In the following lemma we recall the upper {bounds} of the ZCZ width for various types of binary ZCPs.

\begin{lemma}\label{Bound Z-optimal-I}
  Let $(\mathbf{c,d})$ be a binary ZCP of length $N$ {with ZCZ of width} $Z$. Then
\begin{itemize}
\item[1)]  $Z\leq (N+1)/2$ if $N$ is odd \cite{Liu14-2};
\item[2)]  $Z\leq N-2$ if $N$ is even and $(\mathbf{c,d})$  is Type-I ZCP \cite{Liu14}; and
\item[3)]  $Z\leq N-1$ if $N$ is even and $(\mathbf{c,d})$  is Type-II ZCP.
\end{itemize}
\end{lemma}

Note that bound 3) in Lemma~\ref{Bound Z-optimal-I} was obtained by exhaustive computer search. Based on the bounds above, we have the following definition on the Z-optimality of ZCPs.


\begin{definition}\label{def-zopt}\cite{Liu14-2,Liu14}
	A binary ZCP is said to be {\em Z-optimal} if  the upper bound of the ZCZ width in Lemma \ref{Bound Z-optimal-I} is achieved with equality.
\end{definition}

The following lemma gives the lower bounds of the aperiodic autocorrelation sum magnitude outside the ZCZ of a Z-optimal (Type-I {and} Type-II) binary ZCP.

\begin{lemma}\label{lem-optimal-ZCP}
  Let $(\mathbf{c,d})$ be a binary ZCP of length $N$ with ZCZ {of width} $Z$. Then we have the following bounds for odd-length ZCPs and even-length ZCPs.
\begin{itemize}
\item[1)] \cite{Liu14-2} If $(\mathbf{c,d})$ is a Z-optimal Type-I OB-ZCP, then
\begin{equation*}
	|\rho_\mathbf{c}(\tau)+\rho_\mathbf{d}(\tau)|\geq 2,~~~\hbox{for any }(N+1)/2\leq\tau\leq N-1.
\end{equation*}
\item[2)] \cite{Liu14} If $(\mathbf{c,d})$ is a Z-optimal Type-II OB-ZCP, then
\begin{equation*}
	|\rho_\mathbf{c}(\tau)+\rho_\mathbf{d}(\tau)|\geq 2,~~~\hbox{for any }1\leq\tau\leq (N-1)/2.
\end{equation*}
\item[3)] \cite{Liu14} If $(\mathbf{c,d})$ is a Type-I EB-ZCP and $Z\geq N/2$, then
\begin{equation*}
	|\rho_\mathbf{c}(Z)+\rho_\mathbf{d}(Z)|\geq 4.
\end{equation*}
\item[4)] If $(\mathbf{c,d})$ is a Z-optimal Type-II EB-ZCP, then
\begin{equation*}
	|\rho_\mathbf{c}(1)+\rho_\mathbf{d}(1)|\geq 4.
\end{equation*}
\end{itemize}
\end{lemma}

{
The lower bounds above define the optimality of ZCPs, which achieve the smallest possible sum magnitude outside the ZCZ of Z-optimal binary ZCPs.

\begin{definition}\label{def-opt}
  {A Z-optimal Type-I OB-ZCP is called {\em optimal} if the lower bound 1) in Lemma~\ref{lem-optimal-ZCP} is met with equality.}
  And a binary Z-optimal {Type-II} ZCP is called {\em optimal} if the lower bound {2) or 4)} in Lemma~\ref{lem-optimal-ZCP} is met with equality.
\end{definition}
}

 {
\begin{remark}\label{rem-opt}
The bounds in Lemma \ref{Bound Z-optimal-I} may not be tight for all sequence length $N$. For example, as pointed out by one of the anonymous reviewers, there is no Type-I EB-ZCPs with length larger than 14 reported in the literature which can satisfy the upper bound in Lemma \ref{Bound Z-optimal-I}. Therefore it would be possible to derive tigher bounds on ZCZ widths of ZCPs for certain sequence lengths. In such cases, the bounds in Lemma \ref{lem-optimal-ZCP} could be improved as well.
Accordingly, the optimality in Definitions \ref{def-zopt} and \ref{def-opt} should be changed with respect to the new bounds.
\end{remark}
}


\section{A General Construction of Binary ZCPs}\label{section 3}

Infinite families of nontrivial (Type-I and Type-II) OB-ZCPs and Type-I EB-ZCPs were obtained in \cite{Liu14-2} and \cite{Liu14,Chen17,Adhikary18,Xie18}, respectively.  However, {there was no infinite family of Type-II EB-ZCPs in the literature.}
In this section, we present a construction of ZCPs which can generate infinite families of Z-optimal Type-II OB-ZCPs and EB-ZCPs. We first present an example to show that an optimal Type-II EB-ZCP does exist.

\begin{example}\label{Example 1}
Suppose that	
	\begin{equation}
	\begin{split}
	\mathbf{c}&=(+-+++++-+++--+),\\
	\mathbf{d}&=(+-++++-----++-).
	\end{split}
	\end{equation}
	Then $\mathbf{(c,d)}$ is an optimal Type-II EB-ZCP since it is {easily verified} that
	\begin{equation}
	\left|\rho_\mathbf{c}(\tau)+\rho_\mathbf{d}(\tau)\right|_{\tau=0}^{13} =(28,4,0,0,0,0,0,0,0,0,0,0,0,0).
	\end{equation}
\end{example}

Now, we present a systematic construction of Type-II ZCPs in the following.

\begin{construction}\label{Construction}
  Let $\mathbf{a}$ and $\mathbf{b}$ be two binary sequences of length $N$ and $N+1$, respectively. {Taking} $\mathbf{a}$ and $\mathbf{b}$ as seed sequences, {and we then} initialize two sequence pairs $(\mathbf{c}_{0}^1,\mathbf{d}_{0}^1)$ and $(\mathbf{c}_{1}^1,\mathbf{d}_{1}^1)$ as follows:
	\begin{equation}
	\begin{split}
	\mathbf{c}_{0}^1=\mathbf{a}\mid \mid \mathbf{b},&  ~\mathbf{d}_{0}^1=\mathbf{a}\mid \mid -\mathbf{b}; \\
	\mathbf{c}_{1}^1=\mathbf{b}\mid \mid \mathbf{a},&  ~\mathbf{d}_{1}^1=\mathbf{b}\mid \mid -\mathbf{a}. \\
	\end{split}
	\end{equation}
At the $k$-th iteration, we have $2^k$ pairs $(\mathbf{c}_{i}^k,\mathbf{d}_{i}^k)$ for $0\leq i \leq 2^k-1$, given by
\begin{equation}
\begin{split}
\mathbf{c}_{i}^k=\begin{cases}
\mathbf{c}_{\lfloor\frac{i}{2}\rfloor}^{k-1} \mid \mid \mathbf{d}_{\lfloor\frac{i}{2}\rfloor}^{k-1} & \text{ for }i \text{ even},\\
\mathbf{d}_{\lfloor\frac{i}{2}\rfloor}^{k-1} \mid \mid \mathbf{c}_{\lfloor\frac{i}{2}\rfloor}^{k-1} & \text{ for }i \text{ odd}.
\end{cases}
\end{split}
\end{equation}
and
\begin{equation}
\begin{split}
\mathbf{d}_{i}^k=\begin{cases}
\mathbf{c}_{\lfloor\frac{i}{2}\rfloor}^{k-1} \mid \mid -\mathbf{d}_{\lfloor\frac{i}{2}\rfloor}^{k-1} & \text{ for }i \text{ even},\\
\mathbf{d}_{\lfloor\frac{i}{2}\rfloor}^{k-1} \mid \mid -\mathbf{c}_{\lfloor\frac{i}{2}\rfloor}^{k-1} & \text{ for }i \text{ odd}.
\end{cases}
\end{split}
\end{equation}
\end{construction}

\begin{remark}
  Figure \ref{figure tree} illustrates how Construction \ref{Construction} generates {sequence pairs recursively}.
	\begin{center}
		\begin{figure*}
			\centering
			\includegraphics[width=0.8\textwidth]{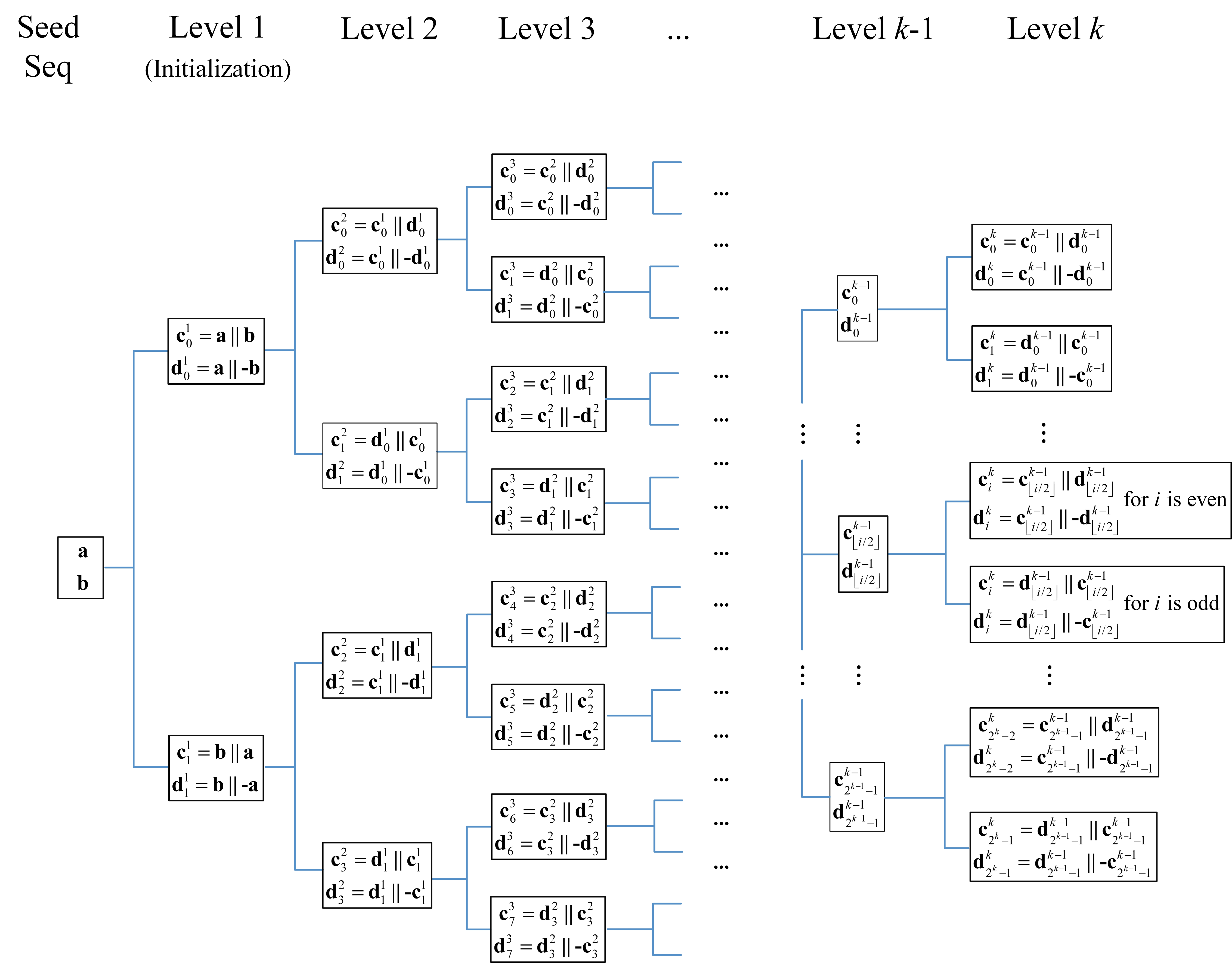}\\
			\caption{Tree representation of recursive Type-II ZCP construction.}\label{figure tree}
		\end{figure*}
	\end{center}
\end{remark}

Based on the construction above, we have the following theorems to obtain {Z-optimal (optimal)} OB-ZCPs and EB-ZCPs.

\begin{theorem}\label{Theorem 1}
Let $\mathbf{a}$ and $\mathbf{b}$ be binary sequences of lengths $N$ and $N+1$, respectively. By Construction \ref{Construction}, the sequence pairs at the initial step, $(\mathbf{c}_0^1,\mathbf{d}_0^1)$ and $(\mathbf{c}_1^1,\mathbf{d}_1^1)$ are Z-optimal Type-II OB-ZCPs of length $L=2N+1$ with ZCZ of width $Z=N+1$.
In addition, {for $r= 0,1$, we have}
\begin{equation}\label{equation Theorem 1}
  \rho_{\mathbf{c}_r^1}(\tau)+\rho_{\mathbf{d}_r^1}(\tau)=
\left\{
  \begin{array}{ll}
    2(\rho_\mathbf{a}(\tau)+\rho_\mathbf{b}(\tau)) , & {1\leq\tau\leq N,} \\
    0, & {N+1\leq\tau\leq2N.}
  \end{array}
\right.
\end{equation}
\end{theorem}

\begin{IEEEproof}
When $r=0$, it is easy to see that
\begin{equation*}
\rho_{\mathbf{c}_r^1}(\tau)=   
\begin{cases}
\rho_\mathbf{a}(\tau)+\rho_\mathbf{b}(\tau) +\sum\limits_{k=0}^{\tau-1}a_{N-1-k}b_{\tau-1-k}, & {1\leq\tau\leq N,} \\
\sum\limits_{k=0}^{2N-\tau}a_kb_{k+\tau-N}, & {N+1\leq\tau\leq2N,}
\end{cases}
\end{equation*}
and
\begin{equation*}
\rho_{\mathbf{d}_r^1}(\tau)= 
\begin{cases}
\rho_\mathbf{a}(\tau)+\rho_\mathbf{b}(\tau) -\sum\limits_{k=0}^{\tau-1}a_{N-1-k}b_{\tau-1-k}, & {1\leq\tau\leq N,} \\
-\sum\limits_{k=0}^{2N-\tau}a_kb_{k+\tau-N}, & {N+1\leq\tau\leq2N.}
\end{cases}
\end{equation*}
Hence, we have
\begin{equation*}
  \rho_{\mathbf{c}_r^1}(\tau)+\rho_{\mathbf{d}_r^1}(\tau)\\=
\left\{
  \begin{array}{ll}
    2(\rho_\mathbf{a}(\tau)+\rho_\mathbf{b}(\tau)) , & {1\leq\tau\leq N,} \\
    0, & {N+1\leq\tau\leq2N,}
  \end{array}
\right.
\end{equation*}
i.e., {the width of} ZCZ is $N+1$. Similarly, we can prove that (\ref{equation Theorem 1}) holds for $r=1$. This completes the proof.
\end{IEEEproof}

\begin{remark}\label{rem11}
  Let $\mathbf{a}$ and $\mathbf{b}$ {be} binary sequences of lengths $N-1$ and $N$, respectively. Then Theorem \ref{Theorem 1} will {produce} Z-optimal Type-II OB-ZCPs of length $2N-1$.
\end{remark}

\begin{theorem}\label{Theorem 2}
  Let $\mathbf{a}$ and $\mathbf{b}$ be binary sequences of lengths $N$ and $N+1$, respectively. By Construction \ref{Construction}, at the $k$-th step of the iteration, sequence pairs $(\mathbf{c}_{r}^k,\mathbf{d}_{r}^k)$ are  Type-II EB-ZCPs of length $L=2^kN+2^{k-1}$ having {the ZCZ of} width $Z=2^kN+2^{k-1}-N$ when $k\geq2$ and $r=0,1,\ldots,2^k-1$.
      In addition, we have
      \begin{equation}\label{equation Theorem 2}
        \rho_{\mathbf{c}_r^k}(\tau)+\rho_{\mathbf{d}_r^k}(\tau)\\=
        \left\{
          \begin{array}{ll}
            2^k(\rho_\mathbf{a}(\tau)+\rho_\mathbf{b}(\tau)), & 1\leq\tau\leq N, \\
            0, & \hbox{otherwise.}
          \end{array}
        \right.
      \end{equation}
\end{theorem}

\begin{IEEEproof}
When $r=0,k\geq2$, it is easy to see that
\begin{equation*}
\rho_{\mathbf{c}_r^k}(\tau)=  
\begin{cases}
\rho_{\mathbf{c}_r^{k-1}}(\tau)+\rho_{\mathbf{d}_r^{k-1}}(\tau) +\rho_{\mathbf{d}_r^{k-1},\mathbf{c}_r^{k-1}}(2^{k-1}N+2^{k-2}-\tau),\\ \hspace{5.8cm} \text{ for } 1\leq\tau\leq 2^{k-1}N+2^{k-2}-1; \\
\rho_{\mathbf{c}_r^{k-1},\mathbf{d}_r^{k-1}}(\tau-(2^{k-1}N+2^{k-2})),\\ \hspace{4cm} \text{ for } 2^{k-1}N+2^{k-2}\leq\tau\leq 2^{k}N+2^{k-1}-1;
\end{cases}
\end{equation*}
and
\begin{equation*}
\rho_{\mathbf{d}_r^k}(\tau)= 
\begin{cases}
\rho_{\mathbf{c}_r^{k-1}}(\tau)+\rho_{\mathbf{d}_r^{k-1}}(\tau) -\rho_{\mathbf{d}_r^{k-1},\mathbf{c}_r^{k-1}}(2^{k-1}N+2^{k-2}-\tau), \\ \hspace{5.8cm} \text{ for } {1\leq\tau\leq 2^{k-1}N+2^{k-2}-1;} \\
-\rho_{\mathbf{c}_r^{k-1},\mathbf{d}_r^{k-1}}(\tau-(2^{k-1}N+2^{k-2})), \\ \hspace{4cm} \text{ for } {2^{k-1}N+2^{k-2}\leq\tau\leq 2^{k}N+2^{k-1}-1.}
\end{cases}
\end{equation*}
Hence, we have
\begin{equation}\label{equation proof of Theorem 2}
\rho_{\mathbf{c}_r^k}(\tau)+\rho_{\mathbf{d}_r^k}(\tau)= 
\begin{cases}
2(\rho_{\mathbf{c}_r^{k-1}}(\tau)+\rho_{\mathbf{d}_r^{k-1}}(\tau)), \hspace{0.5cm} \text{ for } {1\leq\tau\leq 2^{k-1}N+2^{k-2}-1;} \\
0, \hspace{4.2cm} \text{ for } {2^{k-1}N+2^{k-2}\leq\tau\leq 2^{k}N+2^{k-1}-1.}
\end{cases}
\end{equation}

From (\ref{equation Theorem 1}) and (\ref{equation proof of Theorem 2}), {it follows that}
\begin{equation*}
\rho_{\mathbf{c}_r^k}(\tau)+\rho_{\mathbf{d}_r^k}(\tau)\\=
\left\{
\begin{array}{ll}
2^k(\rho_\mathbf{a}(\tau)+\rho_\mathbf{b}(\tau)), & 1\leq\tau\leq N, \\
0, & \hbox{otherwise.}
\end{array}
\right.
\end{equation*}
This completes the proof.
\end{IEEEproof}

\begin{remark}
  Note that the ZCZ width is independent of {different selections} of the seed {sequences $\mathbf{a}$ and $\mathbf{b}$}.
\end{remark}

In the following, we show some illustrative examples.

\begin{example}
Let $\mathbf{a}=(+ +)$ and $\mathbf{b}=(+ + +)$, then according to Construction \ref{Construction} and Figure  \ref{figure tree}, we have at the initial step,

\begin{equation*}
\begin{split}
\mathbf{c}_0^1&=(+++++),\\
\mathbf{d}_0^1&=(++---).
\end{split}
\end{equation*}
At the second iteration,
\begin{equation*}
\begin{split}
\mathbf{c}_0^2&=(+++++++---),\\
\mathbf{d}_0^2&=(+++++--+++).
\end{split}
\end{equation*}
{And at the third iteration,}
\begin{equation*}
\begin{split}
\mathbf{c}_0^3&=(+++++++---+++++--+++),\\
\mathbf{d}_0^3&=(+++++++--------++---).
\end{split}
\end{equation*}
Clearly, we have
\begin{equation*}
\begin{split}
  \left|\rho_{\mathbf{c}_0^1}(\tau)+\rho_{\mathbf{d}_0^1}(\tau)\right|_{\tau=0}^4 &=(10,6,2,\mathbf{0}_2), \\
\left|\rho_{\mathbf{c}_0^2}(\tau)+\rho_{\mathbf{d}_0^2}(\tau)\right|_{\tau=0}^9 &=(20,12,4,\mathbf{0}_7),\\
\left|\rho_{\mathbf{c}_0^3}(\tau)+\rho_{\mathbf{d}_0^3}(\tau)\right|_{\tau=0}^{19} &=(40,24,8,\mathbf{0}_{17}).
\end{split}
\end{equation*}
Hence, $(\mathbf{c}_0^1,\mathbf{d}_0^1)$ is a Z-optimal Type-II OB-ZCP of length $5$ having a ZCZ {of width} $3$. $(\mathbf{c}_0^2,\mathbf{d}_0^2)$ is a Z-optimal Type-II EB-ZCP of length $10$ having a ZCZ {of width} $8$ and $(\mathbf{c}_0^3,\mathbf{d}_0^3)$ is a Type-II EB-ZCP of length $20$ having a ZCZ {of width} $18$. {It is worth noting that} $(\mathbf{c}_0^2,\mathbf{d}_0^2)$ and $(\mathbf{c}_0^3,\mathbf{d}_0^3)$ have {large ZCZ ratios} of $0.8$ and $0.9$, respectively.
\end{example}

\begin{remark}
Note that Z-optimal Type-II EB-ZCPs of length $L=2^k+2^{k-1}$ with ZCZ width $Z=2^k+2^{k-1}-1$ can be generated by Theorem \ref{Theorem 2} when the length of the seed sequence $\mathbf{a}$ is $1$ and $k\geq2$.
\end{remark}

\begin{example}\label{Example optimal EB-ZCP}
Let $\mathbf{a}=(+)$ and $\mathbf{b}=(+ +)$, then according to Construction \ref{Construction} and Figure  \ref{figure tree}, we have at the initial step,

\begin{equation*}
\begin{split}
\mathbf{c}_0^1&=(+++),\\
\mathbf{d}_0^1&=(+--).
\end{split}
\end{equation*}
At the second iteration,
\begin{equation*}
\begin{split}
\mathbf{c}_0^2&=(++++--),\\
\mathbf{d}_0^2&=(+++-++).
\end{split}
\end{equation*}
At the third iteration,
\begin{equation*}
\begin{split}
\mathbf{c}_0^3&=(++++--+++-++),\\
\mathbf{d}_0^3&=(++++-----+--).
\end{split}
\end{equation*}	
Clearly, we have
\begin{equation*}
\begin{split}
  \left|\rho_{\mathbf{c}_0^1}(\tau)+\rho_{\mathbf{d}_0^1}(\tau)\right|_{\tau=0}^2 &=(6,2,0), \\
\left|\rho_{\mathbf{c}_0^2}(\tau)+\rho_{\mathbf{d}_0^2}(\tau)\right|_{\tau=0}^5 &=(12,4,\mathbf{0}_4),\\
\left|\rho_{\mathbf{c}_0^3}(\tau)+\rho_{\mathbf{d}_0^3}(\tau)\right|_{\tau=0}^{11} &=(24,8,\mathbf{0}_{10}).
\end{split}
\end{equation*}
Hence, $(\mathbf{c}_0^1,\mathbf{d}_0^1)$ is an optimal Type-II OB-ZCP of length $3$ having a ZCZ of width $2$. $(\mathbf{c}_0^2,\mathbf{d}_0^2)$ is an optimal Type-II EB-ZCP of length $6$ having a ZCZ of width $5$ and $(\mathbf{c}_0^3,\mathbf{d}_0^3)$ is a Z-optimal Type-II EB-ZCP of length $12$ having a ZCZ of width $11$.
\end{example}

Example \ref{Example optimal EB-ZCP} gives us a construction of Z-optimal Type-II EB-ZCPs. However, the length of Z-optimal Type-II EB-ZCPs which are constructed through the above method is $2^k+2^{k-1}$, and {this makes the length very limited. By} the following construction inspired by the well-known Turyn's construction, we can obtain Type-II ZCPs of length $(2^k+2^{k-1})10^b26^c$.



\begin{theorem}\label{Turyn}
  Let $(\mathbf{c,d}),(\mathbf{e,f})$ be Type-II ZCPs of length $N_1,N_2$ with ZCZ of width $Z_1,Z_2$, respectively. Then $(\mathbf{u,v})$ is a Type-II ZCP of length $N=N_1N_2$ with ZCZ of width $Z=N_1Z_2-N_1+Z_1$. {Here} $(\mathbf{u,v})$ is given by the following formula:
\begin{equation}
  (\mathbf{u,v})=(\mathbf{e}\otimes\frac{\mathbf{c}+\mathbf{d}}{2} +\overleftarrow{\mathbf{f}}\otimes\frac{\mathbf{d}-\mathbf{c}}{2}, \mathbf{f}\otimes\frac{\mathbf{c}+\mathbf{d}}{2} -\overleftarrow{\mathbf{e}}\otimes\frac{\mathbf{d}-\mathbf{c}}{2}),
\end{equation}
where $\otimes$ denotes the Kronecker product. {In particular, if ($\mathbf{e},\mathbf{f}$) is a GCP then ($\mathbf{u},\mathbf{v}$) is a Type-II ZCP of length $N_1N_2$ with ZCZ of width $Z=N_1N_2-N_1+Z_1$.}
\end{theorem}

\begin{IEEEproof}
  {See Appendix.}
\end{IEEEproof}

\begin{example}
Let $\mathbf{c}=(+++)$ and $\mathbf{d}=(+--)$, then $(\mathbf{c},\mathbf{d})$ is a Type-II ZCP of length $3$ with ZCZ of width $2$.
Let $(\mathbf{e,f})$ be a length-$10$ GCP as follows.

\begin{equation*}
\begin{split}
\mathbf{e}&=(++-+-+--++),\\
\mathbf{f}&=(++-+++++--).
\end{split}
\end{equation*}

By Theorem \ref{Turyn}, we obtain a length-$30$ Z-optimal Type-II ZCP $(\mathbf{u,v})$, {as} shown in (\ref{zcp30}),
\begin{equation}\label{zcp30}
	\begin{split}
	\mathbf{u}&=(++++++---+ ++-+++++-+ +- + ++--+--),\\
	\mathbf{v}&=(++- ++--- + ++---- + +-------++++ + +),
	\end{split}
\end{equation}
since {it is computed that}
\begin{equation*}
\left|\rho_{\mathrm{u}}(\tau)+\rho_{\mathrm{v}}(\tau)\right|_{\tau=0}^{29} =(60,20,\mathbf{0}_{28}).
\end{equation*}
	

\end{example}



\section{New Families of Optimal Type-II OB-ZCPs}\label{section 4}

In this section, we construct new families of optimal Type-II OB-ZCPs.
According to Theorem \ref{Theorem 1}, we have constructed Z-optimal Type-II OB-ZCP of any length. In Theorem \ref{Theorem 1}, (\ref{equation Theorem 1}) shows that the key of construction of optimal Type-II OB-ZCPs is to find a sequence pair $(\mathbf{a,b})$ of length $N$ and $N + 1$ with {low} AACSs. The following theorem gives a lower bound for {the AACSs} of the {above} sequence pair $(\mathbf{a,b})$.

\begin{theorem} \label{Theorem 4}
Let $\mathbf{a}$ and $\mathbf{b}$ be two binary sequences of lengths $N$ and $N+1$, respectively. Then
\begin{equation}
  |\rho_\mathbf{a}(\tau)+\rho_\mathbf{b}(\tau)|\geq 1, \hbox{ for all } 1\leq\tau\leq N.
\end{equation}
\end{theorem}

\begin{IEEEproof}
Clearly, we have
\begin{equation*}
	\rho_\mathbf{a}(\tau)\equiv N-\tau \mod 2, \text{ for all } 1\leq\tau\leq N,
\end{equation*}
and
\begin{equation*}
	\rho_\mathbf{b}(\tau)\equiv N+1-\tau \mod 2, \hbox{ for all } 1\leq\tau\leq N.
\end{equation*}
Therefore,
\begin{equation*}
	|\rho_\mathbf{a}(\tau)+\rho_\mathbf{b}(\tau)|\equiv 1\mod 2, \hbox{ for all } 1\leq\tau\leq N.
\end{equation*}
Hence,
\begin{equation*}
  |\rho_\mathbf{a}(\tau)+\rho_\mathbf{b}(\tau)|\geq 1, \hbox{ for all } 1\leq\tau\leq N.
\end{equation*}
\end{IEEEproof}

By exhaustive computer search, seed sequence pairs up to length $N=24$ which can achieve the bound derived in Theorem \ref{Theorem 4}, are listed Table \ref{table 1}.

	\begin{table}[th]\caption{Some seed sequence pair of Lengths up to 24.}\label{table 1}
	\renewcommand{\arraystretch}{0.6}
	\resizebox{\columnwidth}{!}{
		\begin{tabular}{|c|c|c|}
			\hline
			$N$ & $\left(
			\begin{array}{c}
			\mathbf{a} \\
			\mathbf{b} \\
			\end{array}
			\right)
			$ & $\left|\rho_\mathbf{a}(\tau)+\rho_\mathbf{b}(\tau)\right|_{\tau=1}^{N-1}$ \\ \hline\hline
			
			5 & $\bigg( \begin{matrix} ++++- \\ ++--+- \end{matrix} \bigg)$ & $\mathbf{1}_4$ \\
			\hline
			
			6 & $\bigg( \begin{matrix} +++++- \\ ++--+-+ \end{matrix} \bigg)$ & $\mathbf{1}_5$ \\
			\hline
			
			11 & $\bigg( \begin{matrix} ++++++-+--+ \\ +-+---+++--+ \end{matrix} \bigg)$ & $\mathbf{1}_{10}$ \\
			\hline
			
			12 & $\bigg( \begin{matrix} ++++++-++--+ \\ ++----++-+-+- \end{matrix} \bigg)$ & $\mathbf{1}_{11}$ \\
			\hline
			
			13 & $\bigg( \begin{matrix} +----+++-++-+ \\ ------++--+-+- \end{matrix} \bigg)$ & $\mathbf{1}_{12}$ \\
			\hline
			
			14 & $\bigg( \begin{matrix} --++++---+--+- \\ ++++-+++-+-+--+ \end{matrix} \bigg)$ & $\mathbf{1}_{13}$ \\
			\hline
			
			17 & $\bigg( \begin{matrix} -+-+---+++++++--+ \\ +++--++-++-++++-+- \end{matrix} \bigg)$ & $\mathbf{1}_{16}$ \\
			\hline
			
			18 & $\bigg( \begin{matrix} --++--+-+-+++++--- \\ +++-++++-+-++-++--+ \end{matrix} \bigg)$ & $\mathbf{1}_{17}$ \\
			\hline
			
			21 & $\bigg( \begin{matrix} --+--+-++---+-+++---+ \\ -++---------+++-+-+--+ \end{matrix} \bigg)$ & $\mathbf{1}_{20}$ \\
			\hline
			
			22 & $\bigg( \begin{matrix} +++-+----+---++-+--++- \\ +++-+-----+-+----++--+- \end{matrix} \bigg)$ & $\mathbf{1}_{21}$ \\
			\hline
			
			23 & $\bigg( \begin{matrix} +---+---+---++++-+-++-+ \\ ++----+-++-++-----+---+- \end{matrix} \bigg)$ & $\mathbf{1}_{22}$ \\
			\hline
			
			24 & $\bigg( \begin{matrix} -+-----+-+--+++-+--+---+ \\ -+---+----++-++---+++++-+ \end{matrix} \bigg)$ & $\mathbf{1}_{23}$ \\
			\hline
	\end{tabular}}
\end{table}

\begin{example}\label{ex5}
  Let $\mathbf{a}=(+++++-)$ and $\mathbf{b}=(++--+-+)$. Then  {the sequence pair} $(\mathbf{a,b})$ meets the bound in Theorem \ref{Theorem 4}, i.e.,
\begin{equation*}
	\left|\rho_\mathbf{a}(\tau)+\rho_\mathbf{b}(\tau)\right|_{\tau=0}^{6} =(13,\mathbf{1}_6).
\end{equation*}

Also, let
\begin{equation*}
\begin{split}
\mathbf{c}&=\mathbf{a}\mid \mid\mathbf{b},\\
\mathbf{d}&=\mathbf{a}\mid \mid-\mathbf{b}.
\end{split}
\end{equation*}
According to Theorem \ref{Theorem 1},
 $(\mathbf{c,d})$ is a length-$13$ optimal Type-II OB-ZCP having ZCZ width $7$, because
 \begin{equation*}
 	\left|\rho_\mathbf{c}(\tau)+\rho_\mathbf{d}(\tau)\right|_{\tau=0}^{12} =(26,\mathbf{2}_6,\mathbf{0}_6).
 \end{equation*}
It is important to note that optimal Type-II OB-ZCP of length $13$ has not been previously reported in the literature.
\end{example}

{In what follows, we} obtain optimal Type-II OB-ZCPs of lengths $2N\pm 1$, when $N=2^a10^b26^c$, {and $a, b, c$ are non-negative integers}.

\begin{theorem}\label{Construct TCP1}
  Let $(\mathbf{x,y})$ be a binary GCP of length {$N=2^a 10^b 26^c$} and $(\mathbf{a,b})=(\mathcal{V}(\mathbf{x},0),\mathbf{y})$. {Suppose that}
\begin{equation*}
\begin{split}
\mathbf{c}&=\mathbf{a}\mid \mid\mathbf{b},\\
\mathbf{d}&=\mathbf{a}\mid \mid-\mathbf{b}.
\end{split}
\end{equation*}
Then $(\mathbf{c,d})$ is an optimal Type-II OB-ZCP of length $2N-1$ having ZCZ of width $N$.
\end{theorem}

\begin{IEEEproof}
  By the definition of AACF, we have
{
\begin{equation*}
\begin{split}
 \rho_\mathbf{a}(\tau)= & \sum_{i=0}^{N-2-\tau}a_ia_{i+\tau} \\
= & \sum_{i=1}^{N-1-\tau}x_ix_{i+\tau} \\
= & \left(\sum_{i=0}^{N-1-\tau}x_ix_{i+\tau}\right)-x_0x_\tau \\
= & \rho_\mathbf{x}(\tau)-x_0x_\tau,
\end{split}
\end{equation*}
}
for all $\tau=0,1,\ldots,N-1$. As the sequence pair $(\mathbf{x,y})$ is a GCP, then
\begin{equation*}
\begin{split}
  \rho_\mathbf{a}(\tau)+\rho_\mathbf{b}(\tau) = & \rho_\mathbf{x}(\tau)+\rho_\mathbf{y}(\tau)-x_0x_\tau \\
= & \left\{
\begin{array}{ll}
2N-1, & \tau=0, \\
-x_0x_\tau, & \tau=1,2,\ldots,N-1.
\end{array}
\right.
\end{split}
\end{equation*}
Therefore,  we have $|\rho_\mathbf{a}(\tau)+\rho_\mathbf{b}(\tau)| = 1$ for all $\tau=1,2,\ldots,N-1$.
According to Theorem \ref{Theorem 1}, $(\mathbf{c,d})$ is an optimal Type-II OB-ZCP of length $2N-1$.
\end{IEEEproof}

\begin{theorem}\label{Construct TCP2}
  Let $(\mathbf{x,y})$ be a binary GCP of length {$N=2^a 10^b 26^c$} and
\begin{equation*}
\begin{split}
\mathbf{c}&=\mathbf{a}\mid \mid\mathbf{b},\\
\mathbf{d}&=\mathbf{a}\mid \mid-\mathbf{b}.
\end{split}
\end{equation*}
Then $(\mathbf{c,d})$ is an optimal Type-II OB-ZCP of length $2N+1$ where $(\mathbf{a,b})$ is given by
\begin{equation*}
\begin{split}
\mathbf{a}&=\mathbf{x},\\
\mathbf{b}&=\lambda\mid \mid\mathbf{y},~~~\textrm{where } \lambda\in\{+1,-1\}.
\end{split}
\end{equation*}
\end{theorem}

\begin{IEEEproof}
By the definition of AACF, we have
{
\begin{equation*}
\begin{split}
  \rho_\mathbf{b}(\tau)= & \sum_{i=0}^{N-\tau}b_ib_{i+\tau} \\
= & \left(\sum_{i=1}^{N-\tau}b_ib_{i+\tau}\right)+b_0b_\tau \\
= & \left(\sum_{i=0}^{N-1-\tau}y_iy_{i+\tau}\right)+\lambda y_{\tau-1} \\
= & \rho_\mathbf{y}(\tau)+\lambda y_{\tau-1},
\end{split}
\end{equation*}
for all $\tau=1,2,\ldots,N$. As the sequence pair $(\mathbf{x,y})$ is a GCP, then
\begin{equation*}
\begin{split}
  \rho_\mathbf{a}(\tau)+\rho_\mathbf{b}(\tau) = & \rho_\mathbf{x}(\tau)+\rho_\mathbf{y}(\tau)+\lambda y_{\tau-1} \\
= & \left\{
\begin{array}{ll}
2N+1, & \tau=0, \\
\lambda y_{\tau-1}, & \tau=1,2,\ldots,N.
\end{array}
\right.
\end{split}
\end{equation*}
Therefore, $|\rho_\mathbf{a}(\tau)+\rho_\mathbf{b}(\tau)| = 1$ for all $\tau=1,2,\ldots,N$.
}
According to Theorem \ref{Theorem 1}, $(\mathbf{c,d})$ is an optimal Type-II OB-ZCP of length $2N+1$ having a ZCZ of width $N+1$.
\end{IEEEproof}

{
\begin{remark}
 We can also obtain Type-I OB-ZCPs via Construction \ref{Construction} as follows.
  Let $(\mathbf{x,y})$ be a binary GCP of length $N=2^a 10^b 26^c$ and
  \begin{equation}
  \begin{split}
  \mathbf{a}&=\mathbf{x},\\
  \mathbf{b}&=\mathbf{y}\mid \mid\lambda,
  \mathbf{\hat{b}}=\mathbf{y}\mid \mid-\lambda,~~~\textrm{where } \lambda\in\{+1,-1\}.
  \end{split}
  \end{equation}

  Also let
  \begin{equation}
  \begin{split}
  \mathbf{c}&=\mathbf{a}\mid \mid\mathbf{b},\\
  \mathbf{d}&=\mathbf{a}\mid \mid-\mathbf{\hat{b}}.
  \end{split}
  \end{equation}
  Then $(\mathbf{c,d})$ is an optimal Type-I OB-ZCP of length $2N+1$.

  Although the length of above optimal Type-I OB-ZCP has been reported in \cite{Avik tit,Avik iwsda}, our construction is a new method.
  Besides, it is easy to see that $|\rho_\mathbf{a}(\tau)+\rho_\mathbf{b}(\tau)| =|\rho_\mathbf{a}(\tau)+\rho_\mathbf{\hat{b}}(\tau)| = 1$ for all $\tau=1,2,\ldots,N$.
\end{remark}
}


\begin{remark}
  Compared with the {results in} \cite{Liu14-2, Avik tit}, Theorem \ref{Construct TCP1} can construct optimal Type-II OB-ZCPs with more {flexible} parameters. The lengths of the optimal Type-II OB-ZCPs constructed in Theorem \ref{Construct TCP1} are $2N-1$, where $N=2^{a}10^{b}26^c$. With {Theorem \ref{Turyn}}, these optimal Type-II OB-ZCPs and EB-ZCPs can be used to generate EB-ZCPs with ZCZ {of large width}. For example, we can obtain Z-optimal Type-II EB-ZCPs of length $3N$ or $14N$ using {Theorem \ref{Turyn}}. Theorem \ref{Theorem 2} can generate Type-II EB-ZCPs of flexible lengths having large ZCZ widths. The result of Theorem \ref{Construct TCP2} is similar to the Type-II OB-ZCPs reported in \cite{Avik tit}. In Table \ref{ref2} we {give} a complete list of ``best possible" Type-II binary ZCPs up to length $30$ till {now}, which can be constructed by a systematic construction. Note that the lengths in the bold letter in Table \ref{ref2} can only be generated by our systematic construction.
\end{remark}

\begin{remark}
  {We note that} the seed sequence pairs with {low} AACSs, such as the pair used in Example \ref{ex5}, widely exist. We have verified the existence of all binary seed sequence pairs with low AACSs of lengths up to $24$ by computer search.
Some examples of seed sequence pairs with low AACSs up to $24$, obtained by computer search, are shown in Table \ref{table 1}.
Based on Table \ref{table 1} and Theorems \ref{Construct TCP1} and \ref{Construct TCP2}, we are able to construct optimal Type-II OB-ZCP of length $3$ to $53$.

\end{remark}




\section{PMEPR of the Proposed Type-II ZCPs}\label{section 5}

Sequences with low PMEPR are desirable in multi-carrier communications such as orthogonal frequency division multiplexing
(OFDM) systems. In this section, we shall discuss the PMEPR of the proposed Type-II ZCPs. Before doing this, we give a short introduction to the definition of PMEPR of sequences.

We first define the OFDM signal of a sequence $\mathbf{c}=(c_0,c_1,\ldots,c_{L-1})$ to be the real part of
\begin{equation*}
  S_\mathbf{c}(t)=\sum_{k=0}^{L-1}c_ke^{2\pi j(f_c+k\Delta f)t}, 0\leq t\leq \frac{1}{\Delta f},
\end{equation*}
where $j=\sqrt{-1}$, $f_c$ denotes the carrier frequency and $\Delta f$ is the
subcarrier spacing.  Then, the PMEPR of $\mathbf{c}$ (or its OFDM signal) is
defined by
\begin{equation}
\hbox{PMEPR}(\mathbf{c})=\frac{1}{L}\sup_{0\leq t<\frac{1}{\Delta f}}|S_\mathbf{c}(t)|^2.
\end{equation}
For a pair of sequence pair $(\mathbf{c},\mathbf{d})$, its PMEPR is defined as
\begin{equation}
\hbox{PMEPR}(\mathbf{c},\mathbf{d})=\max\{\hbox{PMEPR}(\mathbf{c}),\hbox{PMEPR}(\mathbf{d})\}.
\end{equation}
It turns out in \cite{Liu14-2} that
\begin{equation}\label{equation PMEPR3}
  \hbox{PMEPR}(\mathbf{c},\mathbf{d})\leq 2+\frac{2}{L}\sum_{\tau=1}^{L-1}|\rho_\mathbf{c}(\tau)+\rho_\mathbf{d}(\tau)|
\end{equation}
which reveals a relationship between the PMEPR and autocorrelation of a sequence pair. Clearly, for a GCP $(\mathbf{c}, \mathbf{d})$, one immediately has $\hbox{PMEPR}(\mathbf{c},\mathbf{d})\leq 2$.  Based on (\ref{equation PMEPR3}), upper bounds on  PMEPR of some known Type-I ZCPs were given (see \cite{Liu14-2} and \cite{Chen19} for example). In the sequel, we shall discuss upper bounds for the PMEPR of the proposed Type-II ZCPs based on (\ref{equation PMEPR3}).

The following result follows directly from (\ref{equation PMEPR3}), (\ref{equation Theorem 1}) and (\ref{equation Theorem 2}).

\begin{theorem}\label{th-papr-con1}
Let $\mathbf{a}$ and $\mathbf{b}$ be binary sequences of lengths $N$ and $N+1$, respectively. Then the PMEPR of Type-II ZCP generated by Construction 1 is upper bounded by
\begin{eqnarray}\label{eqn-paprbound-con1}
2+\frac{4}{2N+1}\sum_{\tau=1}^{N}|\rho_\mathbf{a}(\tau)+\rho_\mathbf{b}(\tau)|.
\end{eqnarray}
\end{theorem}

Theorems \ref{th-papr-con1} tells us the upper bound of the PMEPR of the sequences generated by Construction 1
is determined by the aperiodic autocorrelation sums of the seed sequences $\mathbf{a}$ and $\mathbf{b}$. According to Theorem \ref{Theorem 4}, one has
\begin{eqnarray}\label{eqn-AACFs}
\sum_{\tau=1}^{N}|\rho_\mathbf{a}(\tau)+\rho_\mathbf{b}(\tau)|\geq N.
\end{eqnarray}
We thus have following corollary.
\begin{corollary}\label{cor-1}
Let $\mathbf{a}, \mathbf{b}$ be the seed sequences meeting the bound in (\ref{eqn-AACFs}). Then each Type-II ZCP generated by Construction 1 has PMEPR  upper bounded by $2+\frac{4N}{2N+1}\approx 4$.
\end{corollary}

Corollary \ref{cor-1} means that each optimal Type-II OB-ZCP constructed in Section \ref{section 4} of this paper  has PMEPR upper bounded by 4.

{{
\begin{example}
Let $(\mathbf{c},\mathbf{d})$ be the optimal type-II OB-ZCP of length 13 in Example \ref{ex5}. Note that  $(\mathbf{c},\mathbf{d})$ is constructed from the seed sequence $\mathbf{a}=(+++++-)$ and $\mathbf{b}=(++--+-+)$ meeting the bound in \ref{eqn-AACFs}. It is easy to check that
$\hbox{PMEPR}(\mathbf{c})=2.4276$ and $\hbox{PMEPR}(\mathbf{d})=2.4276$ which verifies the result in Corollary \ref{cor-1}.
\end{example}

\begin{theorem}\label{th-papr-con2}
Let $(\mathbf{c}, \mathbf{d})$ be a Type-II ZCP of length $N_1$ with ZCZ of width $Z_1$, $(\mathbf{e,f})$ a GCP of length $N_2$.
Let $(\mathbf{u,v})$ be the Type-II ZCP  generated from $(\mathbf{c}, \mathbf{d})$ and $(\mathbf{e}, \mathbf{f})$ in Theorem \ref{Turyn}.
Then
\begin{eqnarray}\label{eqn-paprbound-turn1}
\hbox{PMEPR}(\mathbf{u},\mathbf{v})\leq  \mathrm{UB}(\mathbf{c,d})
\end{eqnarray}
where
\begin{eqnarray}\label{eqn-paprbound-turn2}
\mathrm{UB}(\mathbf{c,d})=2+\frac{2}{N_1}\sum_{h=1}^{N_1-Z_1}|\rho_\mathbf{c}(h)+\rho_\mathbf{d}(h)|
\end{eqnarray}
is an upper bound of the PMEPR of $(\mathbf{c,d})$.
\end{theorem}
\begin{IEEEproof}
According to Theorem \ref{Turyn}, $(\mathbf{u,v})$ is a Type-II ZCP of length $N=N_1N_2$ and ZCZ width $Z=N_1(Z_2-1)+Z_1$. We then have
\begin{eqnarray}\label{eq-papr-turn-1}
\sum_{\tau=1}^{N-1}|\rho_\mathbf{u}(\tau)+\rho_\mathbf{v}(\tau)|&=&\sum_{\tau=1}^{N-Z}|\rho_\mathbf{u}(\tau)+\rho_\mathbf{v}(\tau)|=N_2\sum_{h=1}^{N_1-Z_1}|\rho_\mathbf{c}(h)+\rho_\mathbf{d}(h)|
\end{eqnarray}
where the second identity follows from  (\ref{equation Th-Turyn}) and the assumption that  $(\mathbf{e,f})$ is a GCP. This together with (\ref{equation PMEPR3}) further leads to
\begin{align*}
\hbox{PMEPR}(\mathbf{u},\mathbf{v})\leq \mathrm{UB}(\mathbf{c,d}).
\end{align*}
Note that $\mathrm{UB}(\mathbf{c,d})$ in (\ref{eqn-paprbound-turn2}) is an upper bound of the PMPER of $(\mathbf{c,d})$ due to (\ref{equation PMEPR3}). This completes the proof.
\end{IEEEproof}

The following result follows directly from Theorem \ref{th-papr-con2} and Corollary \ref{cor-1}.
\begin{corollary}\label{coro-papr-2}
Let $(\mathbf{e,f})$ be a GCP and $(\mathbf{u,v})$ be the Type-II ZCP  in Theorem \ref{Turyn}. Then we have
\begin{itemize}
\item $\hbox{PMEPR}(\mathbf{u},\mathbf{v})< 4$ when $(\mathbf{c,d})$ is the Type-II ZCP generated by Construction 1 from the seed sequences  $\mathbf{a,b}$ meeting the bound in (\ref{eqn-AACFs});

\item  $\hbox{PMEPR}(\mathbf{u},\mathbf{v})\leq  2+\frac{4}{3}\approx 3.33$ when $\mathbf{c}=(+++), \mathbf{d}=(+--)$; and

\item  $\hbox{PMEPR}(\mathbf{u},\mathbf{v})\leq 2+\frac{8}{14}\approx 2.57$ when $\mathbf{c}=(+-+++++-+++--+), \mathbf{d}=(+-++++-----++-)$.

\end{itemize}

\end{corollary}
\begin{example}
In Table \ref{table-papr}, we list the PMEMR of some ZCPs generated by the construction in Theorem \ref{Turyn}. Herein,
$(\mathbf{u}_{3,N_2},\mathbf{v}_{3,N_2})$ (resp., $(\mathbf{u}_{14,N_2},\mathbf{v}_{14,N_2})$ denotes the Type-II ZCP generated from GCP of
length $N_2$ and ZCP $(\mathbf{c,d})$ given by $\mathbf{c}=(+++), \mathbf{d}=(+--)$ (resp., $\mathbf{c}=(+-+++++-+++--+), \mathbf{d}=(+-++++-----++-)$). It can be seen from the table that the PMEPR of these sequences are very close to the bounds in Corollary \ref{coro-papr-2}.
\begin{table}[ht]\caption{PMEPR of some Type-II ZCPs in Theorem \ref{Turyn}.}\centering\label{table-papr}
\begin{tabular}{|c|c|c|c|c|}
  \hline
  Length of GCP $N_2$ & $\hbox{PMEPR}(\mathbf{u}_{3,N_2})$ & $\hbox{PMEPR}(\mathbf{v}_{3,N_2})$ & $\hbox{PMEPR}(\mathbf{u}_{14,N_2})$ & $\hbox{PMEPR}(\mathbf{v}_{14,N_2})$ \\ \hline
  1 & 3.0000 & 1.6667 & 2.5714 & 2.4119 \\
  2 & 2.8452 & 2.6667 & 2.1373 & 2.5137 \\
  4 & 3.0000 & 1.7387 & 2.5714 & 2.5102 \\
  8 & 3.2545 & 3.0740 & 2.5243 & 2.5456 \\
  10 & 3.3333 & 3.0200& 2.4851 & 2.5570 \\
  16 & 3.0312 & 3.2919& 2.5714 & 2.5102 \\
  20 & 3.1834 & 3.3159& 2.5669 & 2.5549 \\
  26 & 3.2902 & 3.2291& 2.5542 & 2.5545 \\
  32 & 3.3290 & 3.1483& 2.5714 & 2.5373 \\
  40 & 3.3333 & 3.0446& 2.5624 & 2.5570 \\
  52 & 3.3064 & 3.3189& 2.5682 & 2.5549 \\
  64 & 3.2902 & 3.3201& 2.5714 & 2.5689 \\
  80 & 3.2037 & 3.2667& 2.5669 & 2.5695 \\
  \hline
\end{tabular}
\end{table}
\end{example}

}

\begin{remark}
ZCPs with good PMEPR properties can
be regarded as potential alternatives of GCPs in practical applications (see \cite{IEEE} for an application scenario) since they can exist for
more lengths. Note that compared to the systematic constructions of Type-I ZCPs with low PMEPR, available in the literature, Type-II ZCPs are available with more flexible lengths.  For example, for sequence lengths $N\in \{N_1\times N_2:  N_1=5,11,13,14,  N_2=2^a 10^b 26^c, a,b,c>0\}$, no Type-I ZCPs were reported in the literature. According to Theorem \ref{Turyn} and Corollary \ref{coro-papr-2}, Type-II ZCPs with low PMEPR exist for such lengths. Therefore, Type-I ZCPs and Type-II ZCPs are two different ways of providing
potential sequences with flexible lengths and low PMEPR  for practical applications.
\end{remark}

%
%
%
%
%

%
%

%
%

%

\section{Concluding Remarks}\label{section 6}
In this paper, some properties and construction of optimal binary ZCPs are studied. Our motivation is the fact {that all currently known binary GCPs have even-lengths of the form $2^a 10^b 26^c$ only}.
{We target at finding optimal binary sequence pairs of any length, which have the closest correlation property to that of GCPs. More precisely,} we proposed a new method which horizontally concatenates sequences $\mathbf{a}$ and $\mathbf{b}$ of different lengths to construct the optimal binary ZCPs. Note that our construction is generic because in our construction $N$ can be any number.
Based on the new method, we constructed optimal and Z-optimal OB-ZCPs with more flexible parameters.
{The main contributions of this paper are listed in the following:}
\begin{enumerate}
  \item For even length of Type-II binary ZCPs, we proved that {the width} of its ZCZ can achieve $N-1$, {in which case} the ZCP is called  a Z-optimal Type-II EB-ZCP.
      We constructed the optimal Type-II EB-ZCPs of lengths $6\times2^a 10^b 26^c$ and $14\times2^a 10^b 26^c$ through Example \ref{Example 1}, Theorem \ref{Theorem 2} and Theorem \ref{Turyn}, where $a, b, c$ are non-negative integers.

  \item We proposed a new recursive construction of Type-II EB-ZCPs.
      By the construction, we can also generate ZCPs with large ZCZ ratio and flexible parameters.

  \item By horizontally concatenating of sequence pair of different lengths, we constructed optimal Type-II OB-ZCPs of length $2N\pm1$, where $N$ is the Golay number, i.e. $N=2^a 10^b 26^c$.

  \item By horizontally concatenating of sequence pair of different lengths, we constructed Z-optimal Type-II OB-ZCPs of length $2N\pm1$, where $N$ can be any number.

  \item We gave upper bounds for the PMEPR of ZCPs from the proposed constructions. It turns out that our constructions can generated Type-II ZCPs with low PMEPR.

  \item {Our construction can also be extended to obtain optimal Type-I OB-ZCPs.  Although the length of generated Type-I OB-ZCP has been reported before, our construction is a new method.  One of our near future work is to explore how to construct optimal Type-I ZCPs with new lengths from Type-II ones.}
\end{enumerate}

We {conclude the present} paper by proposing the following open questions:
\begin{enumerate}
	\item Are there any systematic constructions of optimal Type-II OB-ZCPs in lengths other than the ones discussed in this paper?
	\item Are there more optimal Type-II EB-ZCPs, except for {the lengths $6$ and $14$}?
\end{enumerate}


%

\section*{Appendix\\ Proof of Theorem \ref{Turyn}}
We need the following lemma to prove the theorem.

\begin{lemma}\label{Lem reverse}
	Let $(\mathbf{c,d})$ be a Type-II ZCP of length $N$ with ZCZ of width $Z$, then so is $(\mathbf{c},\overleftarrow{\mathbf{d}})$.
\end{lemma}
\begin{IEEEproof}
	By the definition of AACF, we have
	\begin{equation*}
		\begin{split}
			\rho_{\overleftarrow{\mathbf{d}}}(\tau)= & \sum_{i=0}^{N-1-\tau} \overleftarrow{d_i}\overleftarrow{d_{i+\tau}} \\
			= & \sum_{i=0}^{N-1-\tau} d_{N-1-i}d_{N-1-(i+\tau)} \\
			= & \sum_{t=0}^{N-1-\tau} d_{t+\tau}d_{t} \\
			= & \rho_\mathbf{d}(\tau).
		\end{split}
	\end{equation*}
	Therefore, we have
	\begin{equation*}
		\begin{split}
			\rho_\mathbf{b}(\tau)+\rho_{\overleftarrow{\mathbf{d}}}(\tau) &=\rho_\mathbf{b}(\tau)+\rho_\mathbf{d}(\tau) =0,
		\end{split}
	\end{equation*}
for all $N-Z+1\leq\tau\leq N-1$, and $\rho_\mathbf{b}(N-Z)+\rho_{\overleftarrow{\mathbf{d}}}(N-Z)\neq0$.
\end{IEEEproof}

\subsection*{Proof of  Theorem \ref{Turyn}:}

By the Euclidean division theorem, we have $\tau = k N_{1}+h$ where $0\leq k\leq N_{2}-1, 0\leq h\leq N_1-1$.
By the definition of AACF, we have
{
\begin{equation*}
\begin{split}
\rho_\mathbf{u}(\tau)= & \sum_{m=0}^{N_{2}-1-k} \left[\left(\frac{e_{m}+\overleftarrow{f}_{m}}{2}\right) \left(\frac{e_{m+k}+\overleftarrow{f}_{m+k}}{2}\right) \rho_{\mathbf{c}}(h) \right. + \left(\frac{e_{m}-\overleftarrow{f}_{m}}{2}\right) \left(\frac{e_{m+k}-\overleftarrow{f}_{m+k}}{2}\right) \rho_{\mathbf{d}}(h)\\
&+ \left(\frac{e_{m}+\overleftarrow{f}_{m}}{2}\right) \left(\frac{e_{m+k}-\overleftarrow{f}_{m+k}}{2}\right) \rho_{\mathbf{c}, \mathbf{d}}(h) +\left(\frac{e_{m}-\overleftarrow{f}_{m}}{2}\right) \left(\frac{e_{m+k}+\overleftarrow{f}_{m+k}}{2}\right) \rho_{\mathbf{d}, \mathbf{c}}(h) \\
& + \left(\frac{e_{m}+\overleftarrow{f}_{m}}{2}\right) \left(\frac{e_{m+k+1}+\overleftarrow{f}_{m+k+1}}{2}\right) \rho_{\mathbf{c}}(N_{1}-h)\\&   +\left(\frac{e_{m}-\overleftarrow{f}_{m}}{2}\right) \left(\frac{e_{m+k+1}-\overleftarrow{f}_{m+k+1}}{2}\right) \rho_{\mathbf{d}}(N_{1}-h) \\
& + \left(\frac{e_{m}+\overleftarrow{f}_{m}}{2}\right) \left(\frac{e_{m+k+1}-\overleftarrow{f}_{m+k+1}}{2}\right) \rho_{\mathbf{d},\mathbf{c}}(N_{1}-h) \\&
\left.+\left(\frac{e_{m}-\overleftarrow{f}_{m}}{2}\right) \left(\frac{e_{m+k+1}+\overleftarrow{f}_{m+k+1}}{2}\right) \rho_{\mathbf{c}, \mathbf{d}}(N_{1}-h)\right]
\end{split}
\end{equation*}
}
Therefore, we have

\begin{equation*}
\begin{split}
\rho_\mathbf{u}(\tau)= & ~\frac{\rho_{\mathbf{c}}(h)}{4}\left(\rho_\mathbf{e}(k) +\rho_{\overleftarrow{\mathbf{f}}}(k) +\rho_{\mathbf{e},\overleftarrow{\mathbf{f}}}(k) +\rho_{\overleftarrow{\mathbf{f}},\mathbf{e}}(k)\right)\\&
+\frac{\rho_{\mathbf{d}}(h)}{4}\left(\rho_\mathbf{e}(k) +\rho_{\overleftarrow{\mathbf{f}}}(k) -\rho_{\mathbf{e},\overleftarrow{\mathbf{f}}}(k) -\rho_{\overleftarrow{\mathbf{f}},\mathbf{e}}(k)\right) \\
& + \frac{\rho_{\mathbf{c,d}}(h)}{4}\left(\rho_\mathbf{e}(k) -\rho_{\overleftarrow{\mathbf{f}}}(k) -\rho_{\mathbf{e},\overleftarrow{\mathbf{f}}}(k) +\rho_{\overleftarrow{\mathbf{f}},\mathbf{e}}(k)\right)\\&
+\frac{\rho_{\mathbf{d,c}}(h)}{4}\left(\rho_\mathbf{e}(k) -\rho_{\overleftarrow{\mathbf{f}}}(k) +\rho_{\mathbf{e},\overleftarrow{\mathbf{f}}}(k) -\rho_{\overleftarrow{\mathbf{f}},\mathbf{e}}(k)\right) \\
& + \frac{\rho_{\mathbf{c}}(N_1-h)}{4}\left(\rho_\mathbf{e}(k+1) +\rho_{\overleftarrow{\mathbf{f}}}(k+1) +\rho_{\mathbf{e},\overleftarrow{\mathbf{f}}}(k+1) +\rho_{\overleftarrow{\mathbf{f}},\mathbf{e}}(k+1)\right) \\
& +\frac{\rho_{\mathbf{d}}(N_1-h)}{4}\left(\rho_\mathbf{e}(k+1) +\rho_{\overleftarrow{\mathbf{f}}}(k+1) -\rho_{\mathbf{e},\overleftarrow{\mathbf{f}}}(k+1) -\rho_{\overleftarrow{\mathbf{f}},\mathbf{e}}(k+1)\right) \\
& + \frac{\rho_{\mathbf{c,d}}(N_1-h)}{4}\left(\rho_\mathbf{e}(k+1) -\rho_{\overleftarrow{\mathbf{f}}}(k+1) -\rho_{\mathbf{e},\overleftarrow{\mathbf{f}}}(k+1) +\rho_{\overleftarrow{\mathbf{f}},\mathbf{e}}(k+1)\right) \\
& +\frac{\rho_{\mathbf{d,c}}(N_1-h)}{4}\left(\rho_\mathbf{e}(k) -\rho_{\overleftarrow{\mathbf{f}}}(k+1) +\rho_{\mathbf{e},\overleftarrow{\mathbf{f}}}(k+1) -\rho_{\overleftarrow{\mathbf{f}},\mathbf{e}}(k+1)\right), \\
\end{split}
\end{equation*}
and
\begin{equation*}
\begin{split}
\rho_\mathbf{v}(\tau)
= & ~\frac{\rho_{\mathbf{c}}(h)}{4}\left(\rho_\mathbf{e}(k) +\rho_{\overleftarrow{\mathbf{f}}}(k) -\rho_{\mathbf{e},\overleftarrow{\mathbf{f}}}(k) -\rho_{\overleftarrow{\mathbf{f}},\mathbf{e}}(k)\right) \\&
+\frac{\rho_{\mathbf{d}}(h)}{4}\left(\rho_\mathbf{e}(k) +\rho_{\overleftarrow{\mathbf{f}}}(k) +\rho_{\mathbf{e},\overleftarrow{\mathbf{f}}}(k) +\rho_{\overleftarrow{\mathbf{f}},\mathbf{e}}(k)\right) \\
& - \frac{\rho_{\mathbf{c,d}}(h)}{4}\left(\rho_\mathbf{e}(k) -\rho_{\overleftarrow{\mathbf{f}}}(k) -\rho_{\mathbf{e},\overleftarrow{\mathbf{f}}}(k) +\rho_{\overleftarrow{\mathbf{f}},\mathbf{e}}(k)\right) \\&
-\frac{\rho_{\mathbf{d,c}}(h)}{4}\left(\rho_\mathbf{e}(k) -\rho_{\overleftarrow{\mathbf{f}}}(k) +\rho_{\mathbf{e},\overleftarrow{\mathbf{f}}}(k) -\rho_{\overleftarrow{\mathbf{f}},\mathbf{e}}(k)\right) \\
& + \frac{\rho_{\mathbf{c}}(N_1-h)}{4}\left(\rho_\mathbf{e}(k+1) +\rho_{\overleftarrow{\mathbf{f}}}(k+1) -\rho_{\mathbf{e},\overleftarrow{\mathbf{f}}}(k+1) -\rho_{\overleftarrow{\mathbf{f}},\mathbf{e}}(k+1)\right) \\
& +\frac{\rho_{\mathbf{d}}(N_1-h)}{4}\left(\rho_\mathbf{e}(k+1) +\rho_{\overleftarrow{\mathbf{f}}}(k+1) +\rho_{\mathbf{e},\overleftarrow{\mathbf{f}}}(k+1) +\rho_{\overleftarrow{\mathbf{f}},\mathbf{e}}(k+1)\right) \\
& - \frac{\rho_{\mathbf{c,d}}(N_1-h)}{4}\left(\rho_\mathbf{e}(k+1) -\rho_{\overleftarrow{\mathbf{f}}}(k+1) -\rho_{\mathbf{e},\overleftarrow{\mathbf{f}}}(k+1) +\rho_{\overleftarrow{\mathbf{f}},\mathbf{e}}(k+1)\right) \\
& - \frac{\rho_{\mathbf{d,c}}(N_1-h)}{4}\left(\rho_\mathbf{e}(k) -\rho_{\overleftarrow{\mathbf{f}}}(k+1) +\rho_{\mathbf{e},\overleftarrow{\mathbf{f}}}(k+1) -\rho_{\overleftarrow{\mathbf{f}},\mathbf{e}}(k+1)\right).
\end{split}
\end{equation*}

Therefore, we have
\begin{eqnarray}\label{equation Th-Turyn}
  \lefteqn{\rho_\mathbf{u}(\tau)+\rho_\mathbf{v}(\tau)} \nonumber \\
  & = & ~\frac{\rho_{\mathbf{c}}(h)}{2}\left(\rho_\mathbf{e}(k) +\rho_{\overleftarrow{\mathbf{f}}}(k)\right)
+\frac{\rho_{\mathbf{d}}(h)}{2}\left(\rho_\mathbf{e}(k) +\rho_{\overleftarrow{\mathbf{f}}}(k)\right) \nonumber \\
&  & ~ + \frac{\rho_{\mathbf{c}}(N_1-h)}{2}\left(\rho_\mathbf{e}(k+1) +\rho_{\overleftarrow{\mathbf{f}}}(k+1)\right) +\frac{\rho_{\mathbf{d}}(N_1-h)}{2}\left(\rho_\mathbf{e}(k+1) +\rho_{\overleftarrow{\mathbf{f}}}(k+1)\right) \nonumber \\
& = & ~\frac{1}{2}\bigg[ \big(\rho_{\mathbf{c}}(h)+\rho_{\mathbf{d}}(h)\big) \big(\rho_\mathbf{e}(k) +\rho_{\overleftarrow{\mathbf{f}}}(k)\big) +\big(\rho_{\mathbf{c}}(N_1-h)+ \nonumber \\
&  & ~ \rho_{\mathbf{d}}(N_1-h)\big) \big(\rho_\mathbf{e}(k+1)+\rho_{\overleftarrow{\mathbf{f}}}(k+1)\big) \bigg].
\end{eqnarray}
This together with the definition of Type-II ZCP and Lemma \ref{Lem reverse} means that
\begin{itemize}
\item $\rho_\mathbf{u}(\tau)+\rho_\mathbf{v}(\tau)\neq0$ for $\tau=(N_2-Z_2)N_{1}+N_1-Z_1$ (i.e., $k=N_2-Z_2$ and $h=N_1-Z_1$); and

\item $\rho_\mathbf{u}(\tau)+\rho_\mathbf{v}(\tau)=0$ for all $\tau>(N_2-Z_2)N_{1}+N_1-Z_1$ (i.e., $k>N_2-Z_2$ or ($k=N_2-Z_2$ and $h>N_1-Z_1$)).
\end{itemize}
Therefore, the ZCZ width of $(\mathbf{u}, \mathbf{v})$ is $Z=N_1(Z_2-1)+Z_1$.
This completes the proof of the theorem.


\end{document}